\documentclass[12pt,preprint]{aastex} 

\newcommand{\halpha}{\mbox{H\hspace{0.2ex}$\alpha$}} 
\newcommand{\fei}{Fe\,{\small{I}}}

\shorttitle{quiet Sun mottles} 
\shortauthors{Rouppe van der Voort et al.}

\begin{document}

\title{Magnetoacoustic shocks as driver of quiet Sun mottles}

\author{L. H. M. Rouppe van der Voort\altaffilmark{1,2}}
\email{rouppe@astro.uio.no}

\author{B. De Pontieu\altaffilmark{3}}

\author{V.H. Hansteen\altaffilmark{1,2}}

\author{M. Carlsson\altaffilmark{1,2}}

\author{M. van Noort\altaffilmark{4}} 

\altaffiltext{1}{Institute of Theoretical Astrophysics, University of Oslo,
  P.O. Box 1029 Blindern, N--0315 Oslo, Norway}
\altaffiltext{2}{Center of Mathematics for Applications, University of Oslo,
  P.O. Box 1053 Blindern, N--0316 Oslo, Norway}
\altaffiltext{3}{Lockheed Martin Solar and Astrophysics Lab, 3251 Hanover
  Street, Org. ADBS, building 252, Palo Alto, CA 94304}
\altaffiltext{4}{Institute for Solar Physics of the Royal Swedish Academy of
  Sciences, AlbaNova University Center, SE--106\,91 Stockholm, Sweden}

\begin{abstract}
  We present high spatial and high temporal resolution
  observations of the quiet Sun in \halpha\ obtained with the Swedish 1-m
  Solar Telescope on La Palma.
  We observe that many mottles, jet-like features in the quiet Sun,
  display clear up- and downward motions along their main axis.
  In addition, many mottles show vigorous transverse displacements. 
  Unique identification of the mottles throughout their lifetime is
  much harder than for their active region counterpart, dynamic
  fibrils. 
  This is because many seem to lack a sharply defined edge at their
  top, and significant fading often occurs throughout their lifetime.
  For those mottles that can be reliably tracked, we find that the
  mottle tops often undergo parabolic paths. 
  We find a linear correlation between the deceleration these mottles
  undergo and the maximum velocity they reach, similar to what was
  found earlier for dynamic fibrils.
  Combined with an analysis of oscillatory properties, we conclude
  that at least part of the quiet Sun mottles are driven by
  magnetoacoustic shocks. 
  In addition, the mixed polarity environment and vigorous dynamics
  suggest that reconnection may play a significant role in the
  formation of some quiet Sun jets.
\end{abstract}


\section{Introduction}
\label{sec:intro}

The solar chromosphere in magnetically enhanced regions is dominated
by jet-like features. 
At the limb, they are referred to as spicules, in active region plage
as dynamic fibrils (DFs), and as mottles in the quiet Sun, where they
appear in close vicinity of small flux concentrations in the magnetic
network. 
No consensus has been reached on how these phenomena are related
\citep{grossmann-doerth92spicules_revisited, sterling00spicules}
although it has been suspected for a long time that dark mottles
observed on the disk are the counterparts of the limb spicules, and
that the driving mechanism of these jets have the same origin
\citep{tsiropoula94finestructures, suematsu95disk_spicules,
  christopoulou01finestructure}.

One of the main obstacles for a solid interpretation has been the
difficulty to observationally resolve these less than 1\arcsec\ wide
and short-lived features (3--10~min) that reach heights between 3 and
10~Mm, with quiet Sun jets typically higher than those in active
regions.
Major progress has been made through the use of advanced image
processing in combination with adaptive optics on the Swedish 1-m
Solar Telescope.
Near diffraction-limited ($\sim$120~km) time series in \halpha\ at
extremely high temporal resolution (1~s) have clearly resolved the
dominant temporal and spatial evolution of DFs \citep{hansteen06DFs,
  depontieu07DFs}.

These time series were compared to advanced radiative MHD simulations
and it was demonstrated that the formation of magneto-acoustic shocks
drives the dynamical evolution of fibrils.
The simulations of \citet{hansteen06DFs} and \citet{depontieu07DFs}
expand on the work of \citet{depontieu04nature} who proposed that
photospheric oscillations and convective flows can leak into the
chromosphere, where they shock and drive jets upwards.

The same observational techniques have been exploited to obtain high
quality time series of quiet Sun mottles. 
The combination of high spatial resolution and high temporal cadence
is instrumental to firmly establish the dynamical properties of
mottles. 
As it turns out, this observational requirement could be even more
crucial for quiet Sun mottles than for active regions DFs.


\section{Quiet Sun observations}
\label{sec:observations}

\begin{figure}
\includegraphics[width=\columnwidth]{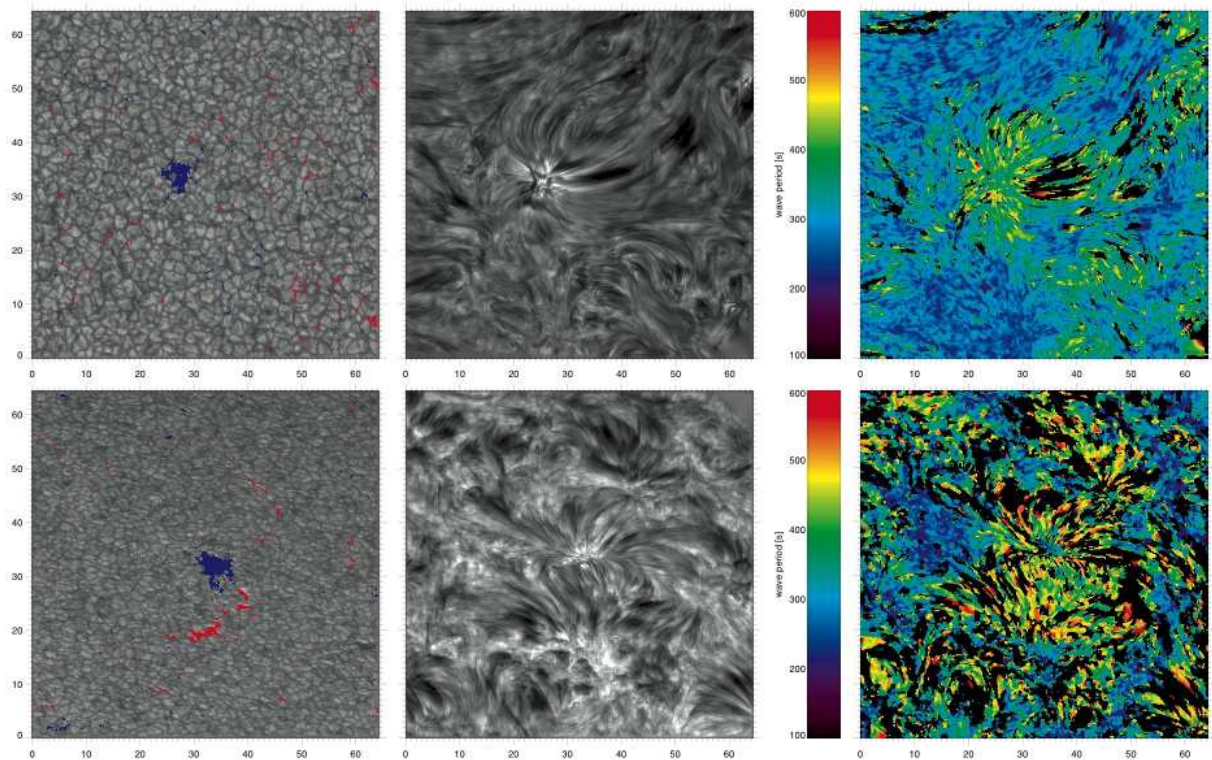}
\caption{The left panels show continuum images, taken closely in time
  with the \halpha\ sequences. Areas in blue and red mark positive and
  negative magnetic flux (absolute magnitude $\ge$250~Mx~cm$^{-2}$ in
  the \fei\,630.2~nm magnetograms). The middle panels show an \halpha\
  line core (top) and a summed \halpha\ $\pm$35~pm (bottom) snapshot
  of the region, whereas the right panels illustrate for each location
  which wave period dominates, i.e., contains the highest number of
  wavepackets with significant power. Units of x and y are in
  arcseconds. Top row data from 18-Jun-2006, bottom row from
  21-Jun-2006.}
\label{fig:oscillations1}
\end{figure}

The observations were obtained with the Swedish 1-m Solar Telescope
\citep[SST,][]{scharmer2003SST} on La Palma, using the adaptive optics
system \citep[AO,][]{scharmer2003AO} in combination with the
Multi-Object Multi-Frame Blind Deconvolution
\citep[MOMFBD,][]{vanNoort05MOMFBD} image restoration method.
The Solar Optical Universal Polarimeter \citep[SOUP,][]{title81SOUP}
provided narrow band images in the \halpha\ line. 
The FOV was about 65\arcsec~$\times$~65\arcsec\ with a pixel scale of
0\farcs65 (the SST diffraction limit at 656.3~nm is 0\farcs165 or
120~km). 
More details on the optical setup and data processing can be found in
\citet{depontieu07DFs}. 
Sets of 37 exposures were included in the restorations, resulting in
several near-diffraction limited \halpha\ line core time series with
1~s cadence: here we analyse a 48~min series from 18-Jun-2006
(observing angle $\theta=20\degr$, or $\mu=\cos \, \theta=0.94$), and
16~min from a 46~min series further towards the limb ($\mu=0.54$) from
21-Jun-2006.
In addition, from the same target area of 21-Jun-2006, we obtained a
54~min Doppler series, with SOUP alternating between the blue and red
wings at $\pm35$~pm from the line core at a cadence of 18.4~s.
Both target areas covered quiet Sun, each featuring at least one
prominent rosette structure that hosts several dark mottles at any
moment. Such rosettes are associated with magnetic network that
consists of strong flux concentrations in the photosphere (see
\halpha\ sample images in Fig.~\ref{fig:oscillations1}).

Before and after the \halpha\ sequences, \fei~630.2~nm Stokes V
magnetograms were recorded with SOUP to provide context information
about the magnetic field topology.
The predominantly mixed polarity magnetic fields (see left panels of
Fig.~\ref{fig:oscillations1}) lead to a very diverse appearance of
dark and bright features in both regions, which seem to have more
complex structuring than in active regions.
Active regions are dominated by both long, horizontal and relatively stable
fibrils, and short dynamic fibrils. 
The quiet Sun has equivalent features: long horizontal dark mottles
and short dynamic mottles, both connecting to the stronger field
concentrations in the network.
In addition, quiet Sun shows many short, highly curved and highly
dynamic features that do not seem to be associated with network, but
mostly appear in the internetwork. 
An example of such region can be seen in the lower right part of
top-center panel of Fig.~\ref{fig:oscillations1}. 
These internetwork features can often be seen underneath the
canopy-like long horizontal mottles, so they seem to be formed at
lower heights. 
This suggests that their dynamics might be associated with the
underlying granular dynamics. 

\section{Analysis and results}
\label{sec:results}

\begin{figure}
  \centering
  \includegraphics[ width=0.49\columnwidth]{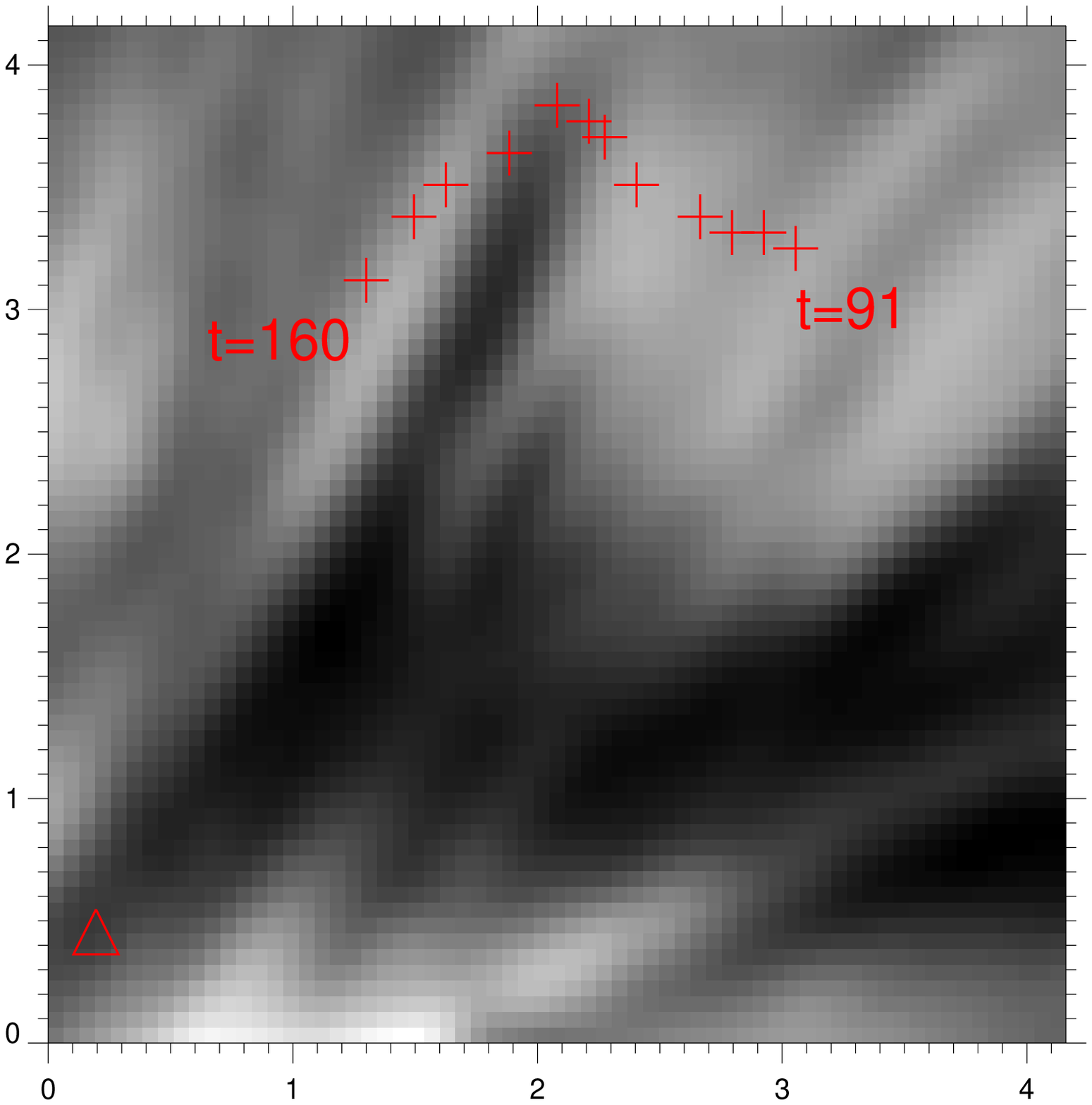}
  \includegraphics[ width=0.49\columnwidth]{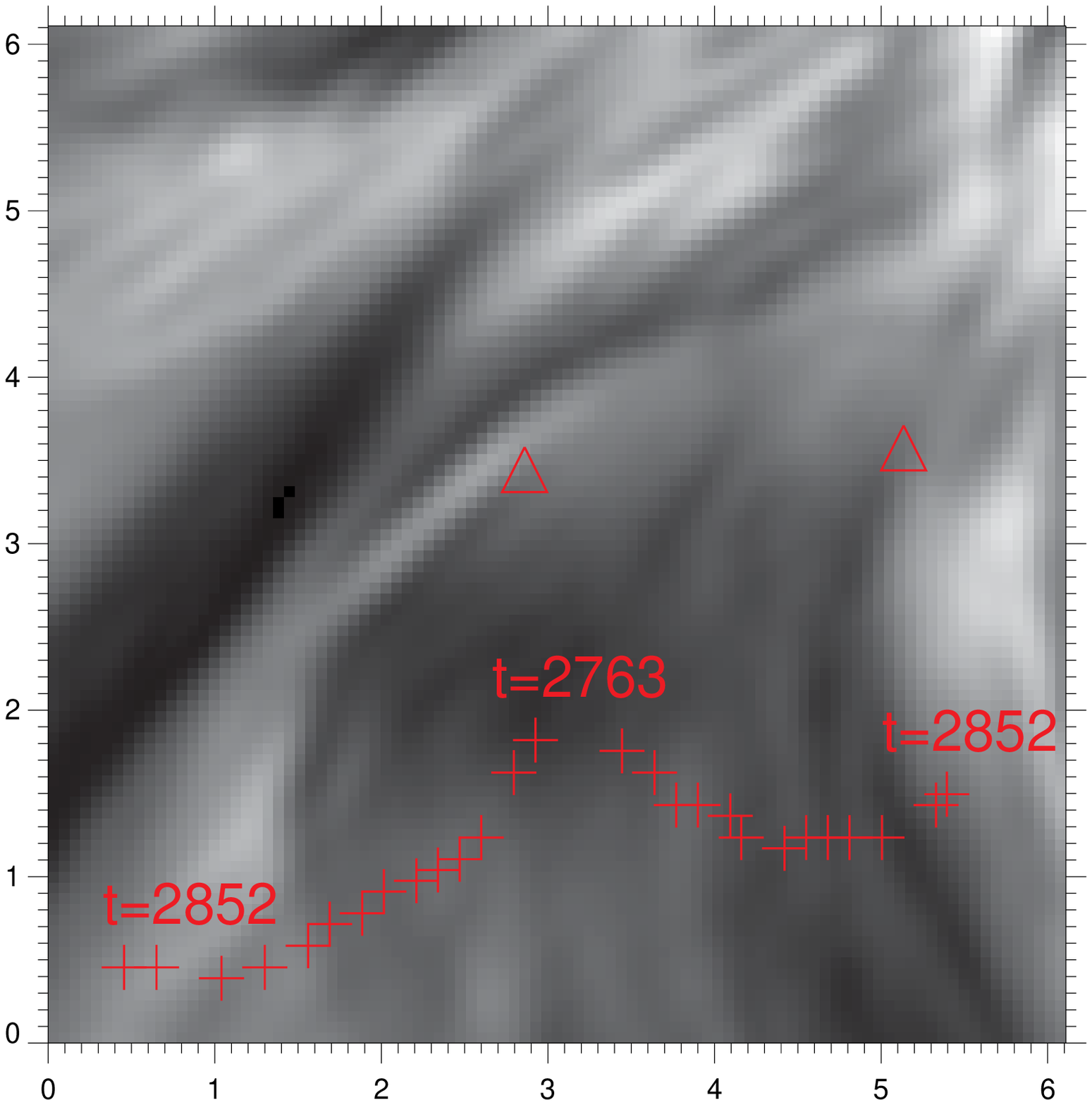}
  \caption{Trajectories of the top end of several dark mottles in
    \halpha\ line core.  Red crosses in the sample images mark the
    manual tracing of the mottle tops. Start and end times are given
    in seconds from the start of the time series. Triangles mark the
    mottle ``roots''. The mottles are rooted in the rosettes in the
    center of the FOV of Fig.~\ref{fig:oscillations1}. The right panel
    shows the paths of 2 mottle tops, moving in opposite
    directions. Note the significant curvature of the mottles in the
    right panel. Units of x and y are in arcseconds.}
  \label{fig:trajectory}
\end{figure}


\begin{figure}
  \centering
  \includegraphics[width=\columnwidth]{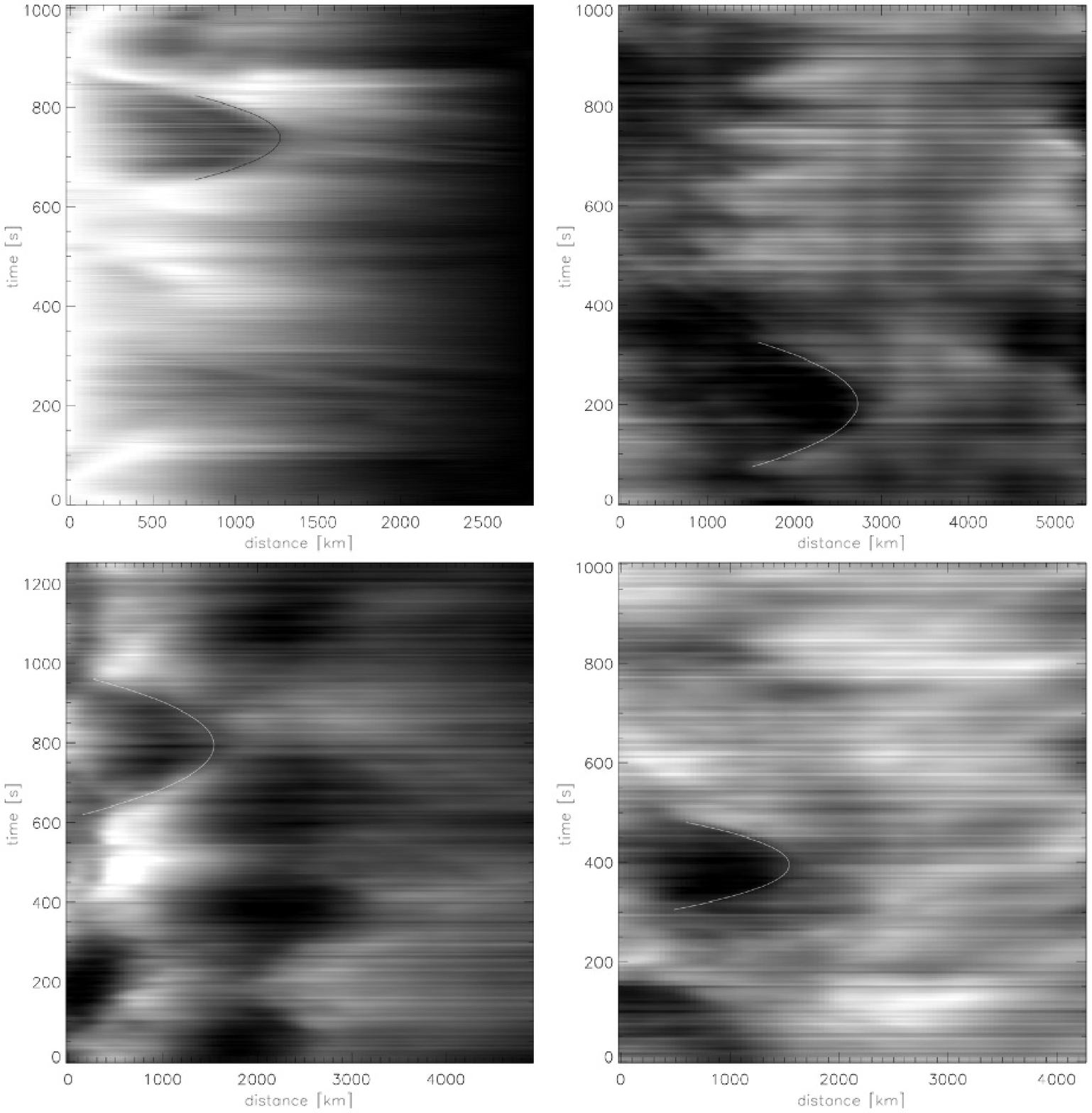}
  \caption{Space-time ('xt') plots of mottles in four different areas
    in \halpha\ line core. To mitigate the effects of transverse
    motion, the xt-plot has been averaged over 10 pixels (0.65\arcsec)
    in the direction perpendicular to the mottle axis.}
  \label{fig:parabolas}
\end{figure}

It is generally more difficult to track quiet Sun mottles during their
lifetime than dynamic fibrils, their active region counterparts.
An important reason for this difficulty is that many mottles lack the
sharply defined edge at the ``top'' end that characterizes the
appearance of DFs in active regions.
In addition, during their lifetime, some mottles display significant
fading that appears to be induced by changing opacity.
Furthermore, most mottles undergo not only up and downward motions
along the direction of the magnetic field, but also significant
transverse motions. 
All these factors often contribute to line-of-sight superposition,
which renders unique identification throughout the lifetime of the
mottle quite challenging.

Figure~\ref{fig:trajectory} shows some examples of the motion of dark
mottles.
In the left panel, the speed transverse to the mottle axis exceeds
20~km~s$^{-1}$ and the transverse displacement is over 2\arcsec\
during 70~s.
The 2 mottles in the right panel display similar speeds and
displacements. 
Many mottles generally undergo some transverse motion, with typical
velocities between 5 and 30~km~s$^{-1}$, although larger apparent
velocities are also present.
We should also note that visual inspection of movies indicates that
what appears as a transverse motion may sometimes be an artifact of
complicated radiative transfer effects combined with coherent
wave-driven motion with phase delay between neighboring (parallel) field
lines. 
Such coherence and phase delays have also been observed in active
region fibrils \citep{hansteen06DFs, depontieu07DFs}.

\begin{figure}
\includegraphics[bb=23 9 456 323, clip, width=\columnwidth]{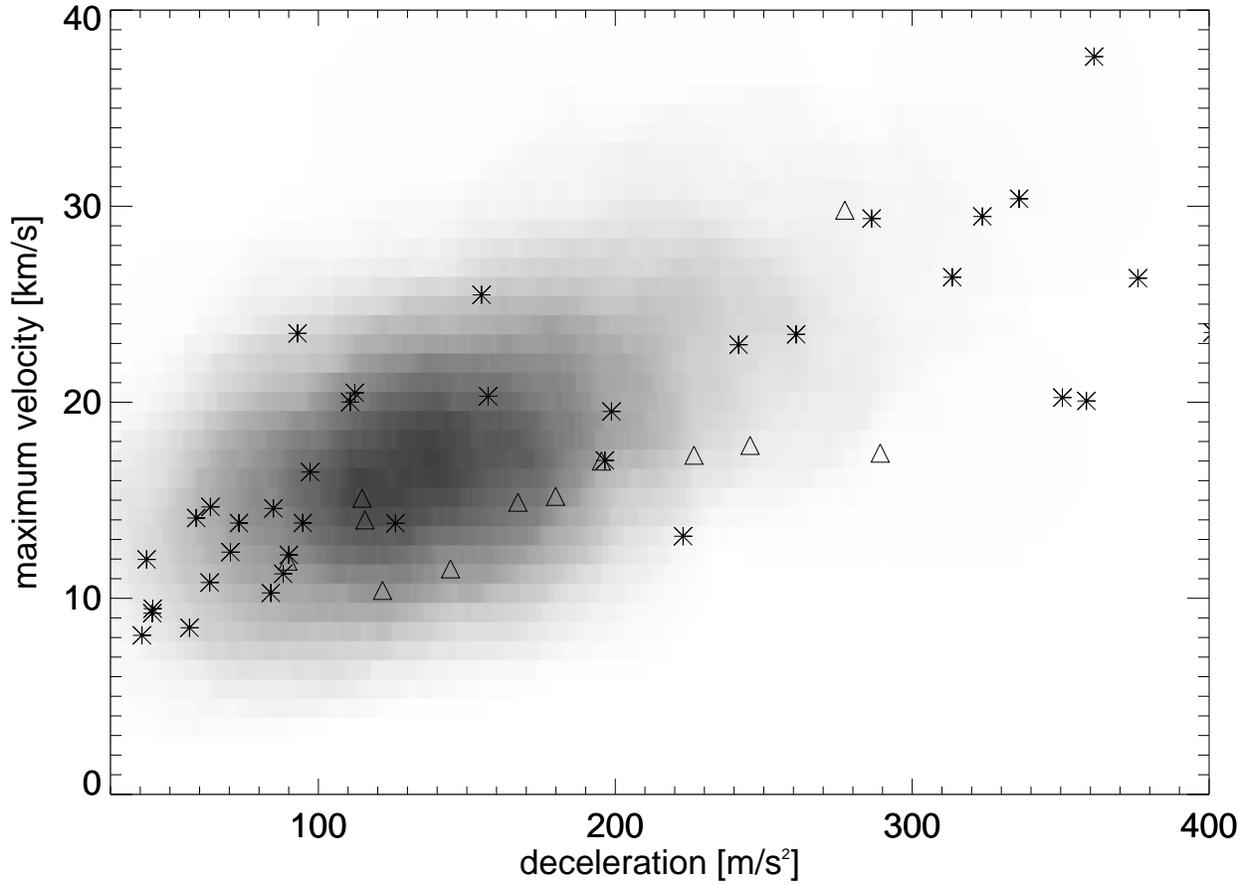}
\caption{Scatterplot of maximum velocity versus deceleration for quiet
  Sun mottles: 37 from 18-Jun-2006 (asterisks), and 12 from
  21-Jun-2006 (triangles). The uncertainty in the measurements is
  estimated to be on the order of 10\%. Shown as an inverse
  grey-scaled density image in the background are the data points for
  DFs from \citet[][see their Fig.~12]{depontieu07DFs}. No correction
  for projection effects have been applied to the data. }
\label{fig:correlations}
\end{figure}

Despite the general difficulty with tracing individual mottles, some
mottles have a sharply-defined top end and do not show as much
transverse motion, so that space-time cuts along the mottle axis can
be extracted.
In such cases, we find that the mottle top often undergoes a parabolic
path, similar to active region fibrils. Parabolic paths had earlier
been suggested by \citet{suematsu95disk_spicules} who used \halpha\
data with much lower resolution than used here. The unprecedented
temporal and spatial resolution of our data for the first time
resolves the parabolic paths of mottles, as shown in
Fig.~\ref{fig:parabolas}.
Following \citet{hansteen06DFs} and \citet{depontieu07DFs}, we
calculate decelerations and maximum velocities for the parabolic paths
of 49 quiet Sun mottles in our data.
Figure~\ref{fig:correlations} shows that the deceleration and
maximum velocity of these mottles is linearly correlated. 
This correlation and the slope between deceleration and maximum
velocity is similar to the one found for DFs, as can be seen in
Fig.~\ref{fig:correlations} (grey-scale background).
Note that the range of decelerations and maximum velocities extends to
somewhat higher values than for DFs. 
This could be caused by the fact that the viewing geometry of quiet
Sun mottles is more diverse, leading to fewer systematic projection
effects, so that relatively more mottles are oriented perpendicular to
the line-of-sight. 
The similarity of mottles and fibrils in the range of values of
maximum velocity and deceleration, and their linear correlation
strongly suggests that these features have a similar driving
mechanism. Jets with similar properties have also been found in recent
numerical simulations \citep{hansteen06DFs,depontieu07DFs,
  depontieu07coimbra}.  This work showed that the correlation between
deceleration and maximum velocity is a clear sign that these jets are
driven by {\em single} slow-mode magnetoacoustic shocks. The
correlation is a natural consequence of "N"-shaped shocks \citep[see
Fig.~4 of ][]{depontieu07coimbra}.
Such a shock wave driving mechanism is also supported by the
observational finding that there are no mottles (or fibrils) with
maximum velocities lower than 8 km/s, the chromospheric speed of
sound.
Additional support comes from the finding of a number of mottles that
have significantly higher deceleration than solar surface gravity
(274~m\,s$^{-2}$).
The deceleration is not related to gravity, but completely determined
by shock wave physics, i.e., the initial pressure pulse, and the
subsequent pressure gradient \citep{heggland07DFsimulations}.
As a result decelerations larger than solar gravity
can occur for fibrils driven by shock waves with large amplitude and
short periods.

A wavelet analysis of the dominant oscillatory behavior
\citep{torrence98wavelets} was performed on the line-core sequence of
18-Jun-2006 and the inner-wing sequence of 21-Jun-2006.
The map of dominant wave power of the line-core series (upper right
panel of Fig.~\ref{fig:oscillations1}) shows significant wavelet power
for at least two wave periods.
The mottles and loops emanating from the network region around
(25\arcsec, 30\arcsec) are all dominated by oscillatory behaviour with
periods around 5--7 minutes, with some of the longer and lower-lying
loops dominated by periods of up to 10 minutes.
This finding is compatible with the idea that leakage of global
oscillations from the photosphere (with dominant periods around 5
minutes) are important in the formation and dynamics of
network-associated mottles.
The internetwork regions, such as the lower left region, show periods
that are closer to 3 minutes. This suggests that internetwork regions
(where the field is not as dominant) are dominated by waves with
periods around the chromospheric acoustic cutoff period of 3 minutes.
It is interesting to note that the 3 min power is also often visible
in regions where the long, low-lying loops, e.g. around (45\arcsec,
45\arcsec), that are typically dominated by longer periods, start to
become transparent.  This suggests that in such regions lower opacity
of the overlying loops allows glimpses of the internetwork dynamics
and oscillations underneath.

A similar picture emerges from the H$\alpha\,\pm\,35$~pm summed
time series shown in the lower right panel of
Fig.~\ref{fig:oscillations1}. 
The network-associated loops are dominated by periods of 5 min or
longer (e.g., around 35\arcsec, 35\arcsec), whereas the internetwork
regions (e.g., around 15\arcsec, 40\arcsec) are dominated by 3 minute
oscillations. This again strongly suggests that leakage of
photospheric oscillations into the chromosphere dominates much of the
dynamics of the quiet network chromosphere.  There is a hint that
right at the center of the network regions, periods closer to 3
minutes appear to occur more often (e.g., at 33\arcsec,
35\arcsec). This is presumably where the field is more vertical, so
that the acoustic cutoff period reverts to its nominal 3 min value
\citep[similar to the dense plage region in][]{hansteen06DFs}.  We also
see slightly longer periods dominating the loops connecting two
opposite polarity network regions (one centered at 35\arcsec,
32\arcsec, and one at 30\arcsec, 18\arcsec). These longer periods are
reminiscent of the periods that are seen in low-lying long fibrils
that originate in sunspots or strong plage
\citep{depontieu07DFs}. This may not be a coincidence, as the quiet
network regions shown here actually contain a stronger than usual
amount of magnetic flux (some tiny pores are even visible in the
continuum images).
%


\section{Conclusions}
\label{sec:conclusions}

The parabolic paths, correlation between deceleration and maximum
velocity and oscillatory properties of quiet Sun mottles strongly
suggest that the mechanism that drives active region dynamic fibrils
is also responsible for the formation of at least a subset of quiet
Sun mottles. This is not surprising, since all of the ingredients of
the fibril mechanism are also present in quiet Sun: global
oscillations and convective flows that are guided into the
chromosphere along magnetic field lines. How is this mechanism
modified under quiet Sun conditions? The generally weaker magnetic
fields of the quiet Sun imply that the height of the plasma $\beta=1$
surface is generally higher than in active region plage. That means
that the field is less rigid at these heights: chromospheric flows and
waves can influence the motion of magnetic field lines up to larger
heights. A less rigid field would lead to a much more dynamic magnetic
field at upper chromospheric heights, with significantly more
transverse motions. This is exactly what we observe in our quiet Sun
data. Mode coupling between different wave modes at the plasma
$\beta=1$ surface can also be expected to play a large role in this
magnetic environment \citep{bogdan03waves}.  In fact, the 2D
radiative MHD simulations of \citet{hansteen06DFs} and
\citet{depontieu07DFs} clearly show that under weaker field conditions
than found in active region plage, fast magnetoacoustic waves play a
significant role in the dynamics of fibril-like jets.  These fast
modes propagate perpendicular to the fibril-axis, and can lead to a
significant transverse motion of the whole fibril-like jet. More
advanced 3D simulations will be necessary to determine the role of
Alfv\'en waves in explaining transverse motions of mottles.

While we have focused on relatively ``well-behaved'' mottles here, our
observations also show many examples where significant reorganizations
of the magnetic field occur, with apparent (un?)twisting and motions
at Alfv\'enic speeds.  Such reorganizations are most probably signs of
magnetic reconnection caused by the dynamic magneto-convective driving
of mixed polarity fields in the quiet Sun. Given these observations
and the evidence presented in the above for a fibril-like driving
mechanism for quiet Sun mottles, it seems quite probable that both
reconnection and chromospheric shock waves play a role in jet
formation in quiet Sun. The presence of multiple driving mechanisms
may well be the main reason why the spicule problem has been so
difficult to resolve. 


\acknowledgments
This 
research was supported through grants 146467/420
and 159137/V30 of The Research Council of Norway.
B.D.P. was supported by NASA grants NNG06GG79G, NNG04-GC08G and
NAS5-38099 (TRACE)
and would like to thank ITA for excellent hospitality in
August 2006. 
The Swedish 1-m Solar Telescope is operated on the island of La Palma
by the Institute for Solar Physics of the Royal Swedish Academy of
Sciences in the Spanish Observatorio del Roque de los Muchachos of the
Instituto de Astrof{\'\i}sica de Canarias.



\end{document}